\documentclass[twocolumn,showpacs,preprintnumbers,prb,fleqn,floatfix]{revtex4}

\usepackage{graphicx}
\usepackage{dcolumn}
\usepackage{bm}

\begin{document}

\title{Heat transport in $Bi_{2+x}Sr_{2-x}CuO_{6+\delta}$: departure from the Wiedemann-Franz law
in the vicinity of the metal-insulator transition}

\author{Cyril Proust$^{1,2}$, Kamran Behnia$^{1}$, Romain Bel$^{1}$,
Duncan Maude$^{3}$ and S. I. Vedeneev $^{4}$}

\affiliation{$^{1}$Laboratoire de Physique Quantique (CNRS),
 ESPCI, 10 rue Vauquelin, 75231 Paris, France  \\
$^2$ Laboratoire National des Champs Magn\'{e}tiques
Puls\'{e}s (CNRS-UPS-INSA), BP 14245, 31432 Toulouse, France \\
$^3$ Grenoble High Magnetic Field Laboratory (CNRS), BP 166,
38042 Grenoble, France\\
$^4$ P.N. Lebedev Physical Institute, Russian Academy of Sciences,
119991 Moscow, Russia \\ }

\date{\today}

\begin{abstract}
We present a study of heat transport in the cuprate superconductor
$Bi_{2+x}Sr_{2-x}CuO_{6+\delta}$ at subkelvin temperatures and in
magnetic fields as high as 25T. In several samples  with different
doping levels close to optimal, the linear-temperature term of
thermal conductivity was measured both at zero-field and in
presence of a magnetic field strong enough to quench
superconductivity. The zero-field data yields a superconducting
gap of reasonable magnitude displaying a doping dependence similar
to the one reported in other families of cuprate. The normal-state
data together with the results of the resistivity measurements
allows us to test the Wiedemann-Franz(WF) law, the validity of
which was confirmed in an overdoped sample in agreement with
previous studies. In contrast, a systematic deviation from the WF
law was resolved for samples displaying either a lower doping
content or a higher disorder. Thus, in the vicinity of the
metal-insulator cross-over, heat conduction in the
zero-temperature limit appears to become significantly larger than
predicted by the WF law. Possible origins of this observation are
discussed.
\end{abstract}

\pacs{ 74.25.Fy, 74.72.Hs}

\maketitle

\section{Introduction}
The study of subkelvin heat transport in high-T$_c$ cuprates has
been a subject of considerable interest over the past few years.
The pioneer work by Taillefer and co-workers\cite{Taillefer97}
resolved a finite linear term in the thermal conductivity of
superconducting YBa$_2$Cu$_3$0$_{7-\delta}$ (YBCO) in the
zero-temperature limit. Subsequently, this experimental probe has
been the object of a large number of
experimental\cite{Taillefer97,Aubin99,Chiao99,Chiao00,Nakamae01,Takeya02,Sutherland03,Sun04,Ando04}
and theoretical\cite{Sun95,Graf96,kubert98,Durst00}investigations.
It appears that these measurements have opened a new window on
transport by the nodal quasi-particles of the superconducting
state \cite{Hussey02}.

A number of remarkable findings have emerged. In agreement with
theoretical predications for a d-wave
superconductor,\cite{Graf96,Durst00} the zero-temperature thermal
conductivity of optimally-doped cuprates was found to be
\emph{universal},\cite{Taillefer97,Nakamae01} in the sense that a
strong variation of the scattering time of the quasi-particles due
to disorder has little effect on the magnitude of the electronic
thermal conductivity in the zero-temperature limit,
$\kappa_{0}/T$. Moreover, the superconducting gap extracted from
$\kappa_{0}/T$ has a nodal slope and a size comparable to what has
been measured by other techniques.\cite{Chiao00} On the other
hand, $\kappa_{0}/T$ was found to show a strong doping
dependence.\cite{Takeya02,Sutherland03} It decreases steadily as
the Mott insulator is approached. If one assumes that
$\kappa_{0}/T$ continues to inversely scale with the
superconducting gap in the underdoped regime, this result points
to a superconducting origin for the pseudogap.\cite{Sutherland03}
Finally, the field dependence of thermal conductivity has been
another source of insight. In the optimally-doped cuprates,
$\kappa_{0}/T$ was found to increase as a function of magnetic
field,\cite{Aubin99,Chiao99} providing an experimental
confirmation of the  Volovik excitations\cite{Volovik93} expected
in a \textit{d}-wave superconductor. In underdoped
La$_{2-x}$Sr$_x$CuO$_4$ (LSCO), however, $\kappa_{0}/T$
\emph{decreases} with magnetic field\cite{Hawthorn03,Sun03} in the
mixed state and is reminiscent of the metal-insulator transition
observed by resistivity measurements in the normal
state.\cite{Ando95}

In principle, the study of thermal transport in the \emph{normal}
state of the cuprates allows to test what is generally believed to
be a robust signature of a Fermi liquid, namely the
Wiedemann-Franz (WF) law. According to this law, providing that
collisions of electrons are elastic (which is the case in the
zero-temperature limit), thermal and electrical conductivities are
related by a universal constant.

\begin{equation}
\frac{\kappa}{\sigma T}=L_0
\label{WF}
\end{equation}
where $L_0=2.44 \times 10^{-8}~W \Omega K^{-2}$ is Sommerfeld's
value.

One obvious difficulty, in performing such a test, is to attain
the normal state in the T=0 limit.  The magnetic field needed to
destroy the superconducting ground state in archetypal
optimally-doped cuprates is simply too large. Therefore, attempts
to test the validity of the WF law in various cuprates were all
made in compounds where superconductivity is (for one reason or
another) much weaker or absent. The first study, on the
electron-doped Pr$_{2-x}$Ce$_x$CuO$_4$ at optimal doping, led Hill
\emph{et al.} to report a violation of the WF law for the first
time in any metal.\cite{Hill01} The most striking feature of the
data was a vanishing of the electronic thermal conductivity both
in the normal and in the superconducting states below 0.3K which
is now believed to be a consequence of decoupling between electron
and phonon thermal baths.\cite{Smith05} As emphasized by Hill
\emph{et al.}, the electronic thermal conductivity extracted above
0.3~K yielded $\frac{\kappa}{\sigma T} \sim 1.7 L_0$. However, the
presence of the extrinsic downturn hampered the solid
establishment of such a violation of the WF law in the T=0 limit.

In a second study, Proust \emph{et al.} provided compelling
evidences for the validity of the WF law in overdoped Tl-2201 with
an accuracy of a few percent.\cite{Proust02} Finally, Nakamae
\emph{et al.} confirmed this validity in the case of heavily
overdoped \emph{non-superconducting} LSCO.\cite{Nakamae03}
Moreover, the latter study indicated that the downturn of thermal
conductivity reported by Hill \emph{et al.}, is extrinsic as it
was reproduced in a system displaying a Fermi liquid ground state.
The observed validity of the WF law in overdoped cuprates is in
conformity with the general belief that the ground state in this
regime is indeed a Fermi liquid.\cite{Hussey03}

 In this paper, we report on a study of heat transport in
$Bi_{2+x}Sr_{2-x}CuO_{6+\delta}$, a cuprate,  which, even at
optimal doping, presents an unusually low T$_c$ ($\sim$ 10~K) and
a relatively accessible H$_{c2}$. In a previous communication on
the first part of these measurements,\cite{Bel04} we reported that
for an optimally-doped sample, $\kappa/\sigma T=1.3 L_0$. This
small departure from the Wiedemann-Franz law appeared to be
nevertheless significant as it was larger than the experimental
error. Here, we present new data which shows that the departure is
even larger in two other samples and systematically increases with
underdoping and/or disorder. The picture emerging from this study
is a pronounced violation of the WF law in the vicinity of the
metal-to-insulator transition in cuprates. What is observed is a
roughly two-fold excess of heat conductivity as
originally-reported in PCCO by Hill \emph{et al.} for T$>$0.3K.
Our observation remains robust down to the lowest temperatures
explored and is not contaminated by the extrinsic
downturn\cite{Smith05} present in PCCO. Moreover, our study of the
superconducting state in Bi$_2$Sr$_2$Cu0$_{6+\delta}$ is in
conformity with features reported in other cuprates, indicating
that heat transport in this member of the cuprate family does not
fundamentally differ from the others.

\section{The measurement and its precision}

In-plane thermal conductivity was measured with a standard
two-thermometer-one-heater setup which allows to measure the
in-plane electrical resistivity under the same conditions.  A
number of zero-field measurements reported here were performed at
the ESPCI in Paris. However, in order to measure subkelvin thermal
conductivity in a Bitter/polyhelix magnet at the Grenoble High
Magnetic Field Laboratory (GHMFL), we developed a new experimental
set-up.\cite{Linden04} The water-cooled resistive magnet generates
mechanical vibrations, a part of which is inevitably transmitted
to the thermometers coupled to the sample. In order to overcome
this problem, our set-up containing the sample and thermometers is
placed inside a vacuum chamber, which can be introduced into the
mixing chamber of a commercial top-loading dilution refrigerator
with a relatively large cooling power ($\sim 300~\mu$W at
$100$~mK).

The second technical challenge was to accurately measure the
temperature since the magnetoresistance of the ruthenium oxide
thermometers is far from negligible. For this purpose, we used a
coulomb blockage thermometer (CBT), a primary thermometer know to
show no variation with magnetic field.\cite{Pekola94} This
thermometer has been used \emph{in situ} in order to calibrate in
a magnetic field, the RuO$_2$ sensors used to measure local
temperature of the sample.

\begin{figure}
\includegraphics[width=0.9\linewidth,angle=0,clip]{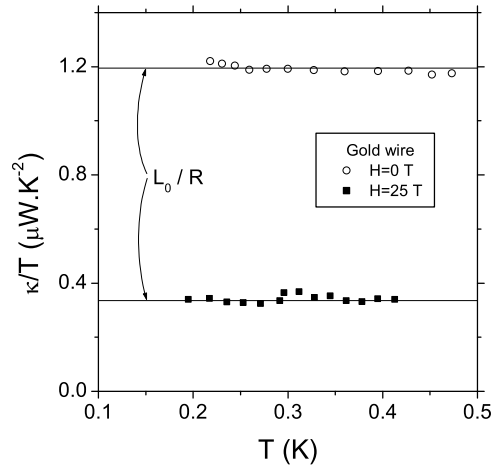}
\caption{Thermal conductance of a gold wire for H=0 (open circle)
and for H=25~T (solid squares) compared to the magnitude expected
by the WF law (solid lines).} \label{fig1}
\end{figure}

In order to estimate the experimental resolution, we have verified
the WF law in a gold wire by measuring thermal conductance and
electrical resistance in zero magnetic field and for H=25~T. As
seen in Fig.~\ref{fig1}, the thermal conductivity is purely
electronic and therefore a linear function of temperature. The
magnitude of the linear term can be compared to what is expected
according to the WF law (solid lines in Fig.~\ref{fig1}). Both at
zero magnetic field and H=25~T, the magnitude of $\kappa/T$ is
very close to L$_0$/R, where R is the measured resistance of the
wire. The application of a magnetic field leads to a threefold
decrease of $\kappa/T$, reflecting the positive magnetoresistance
of gold in this temperature range. The discrepancy between
$\kappa/T$ and L$_0$/R is 1$\%$ (3$\%$) for H=0 (H=25~T), giving
an estimation of experimental precision. Since the electrical
resistance of the gold wire is comparable to the resistance of the
normal state of the cuprate samples measured in our investigation,
this also provides a verification of the absolute value of thermal
conductivity obtained in our study and shows that the thermal leak
of our set-up is negligible.

The major source of uncertainty when determining the absolute
value of $\kappa$ and $\sigma$ is due to the geometric factors
which are about $\pm10~\%$. In principle, the verification of the
WF law should not suffer from this uncertainty, since no geometric
factor enters in Eq.~\ref{WF} and the same contacts are used for
measuring electrical voltage and temperature. In practice, the
geometric factor could be slightly different for electric and
thermal transport, if the width of gold pads evaporated on the
sample is not negligible over the distance between the two voltage
(thermometer) contacts. However, we did not observe any
correlation between the magnitude of thermal conductivity and the
width of gold electrodes. Additional sources of uncertainty come
from an eventual c-axis contamination of the in-plane
conductivity, which would lead to an overestimation of $\rho_0$
and the extrapolation of linear term of the thermal conductivity
at T=0. The sum of the identified sources of experimental error
yields an uncertainty of $\approx20\%$ for the verification of the
WF law.

\section{Samples, doping level and disorder}

We have studied five single crystals of
$Bi_{2+x}Sr_{2-x}CuO_{6+\delta}$. They were grown in a gaseous
phase in closed cavities of a KCl solution melt as detailed
elsewhere.\cite{Gorina94} Typical dimensions of the crystals are
(2-10)$\times $(400-800)$\times $ (600-900) $\mu $m$^{3}$. Four
gold pads are evaporated onto the surface of the samples with a
typical contact resistance of about 1~ohm.

Starting from the insulating phase $Bi_2Sr_2CuO_6$, there are two
possibilities to add carriers in the $CuO_2$ planes. One may
change the amount $\delta$ of excess oxygens in BiO planes or
substitute $Sr^{2+}$ ions by trivalent ions, such as $La^{3+}$ or
excess of $Bi^{3+}$. Due to the difficulty of changing the oxygen
content in a controlled way, $as-grown$ crystals were used in this
study. The variation of T$_c$ is set by changing the $Bi$ and $Sr$
contents.

The maximum $T_c$ found in $Bi_{2+x}Sr_{2-x}CuO_{6+\delta}$ was
found to be $\approx$~10 K in agreement with previous studies.
This is much lower than the T$_c$ of 33~K in the La-doped
($Bi_{2}Sr_{2-x}La_{x}CuO_{6+\delta}$) system. The very low level
of T$_{c}$ in La-free Bi-2201 has been recently addressed by a
careful study of Eisaki \emph{et al.},\cite{Eisaki04} who probed
the impact of  $Sr$ substitution by ions with different radius on
the physical properties of Bi-2201. They suggest that cation
disorder, particularly at the $Sr$ site, strongly  affects the
maximum attainable T$_c$. Based on the ionic radii shown by Ahrens
\cite{Ahrens}, they correlate the size of the ionic radius of
Sr$^{2+}$ ($1.12~\AA$), La$^{3+}$ ($1.14~\AA$) and Bi$^{3+}$
($0.96~\AA$) with the magnitude of T$_{c}$. The mismatch between
the ionic radii of Sr$^{2+}$ and Bi$^{3+}$ causes additional
lattice distortions and may account for the low T$_{c}$ in
$Bi_{2+x}Sr_{2-x}CuO_{6+\delta}$. It is worth to notice that
Shannon\cite{Shannon} found different ionic radius values since it
depends on the coordinates and its surrounding in the lattice.

\begin{figure}
\includegraphics[width=0.9\linewidth,angle=0,clip]{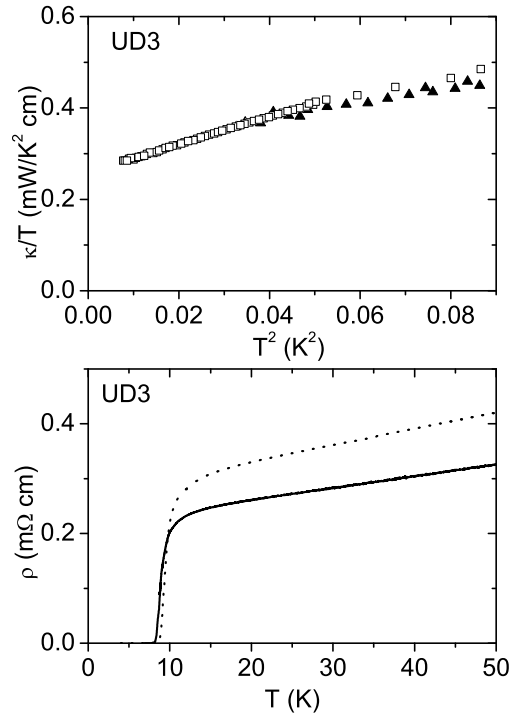}
\caption{Upper panel: Thermal conductivity of sample UD3 measured
a first time (solid triangles) and a second time using another
set-up three months later (open squares). Lower panel: Temperature
dependence of the resistivity of the same sample measured under
the same conditions. Solid (dotted) line is the first (second)
measurement.} \label{fig2}
\end{figure}

Excess oxygen in $BiO$ planes would also engender a random
potential in the CuO$_2$ planes but in a less controlled way.
Indeed, Ono and Ando \cite{Ono03} have already noticed that the
magnitude of the resistivity at T$_c$ can vary from sample to
sample at the same doping level because of inhomogeneous oxygen
concentration in the sample. During our study, we have noticed
that with aging and thermal cycle, our underdoped samples are not
stable and evolve. This is illustrated by the data shown in
Fig.~\ref{fig2}. Bottom panel shows the temperature dependence of
the resistivity of the as-grown sample UD3 (solid line) and during
a second measurement performed a few months after (dotted line).
The corresponding low-temperature thermal conductivity  is
displayed in the top panel of the same figure. Note that during
the first measurement (solid triangles) performed at GHMFL, the
data points stop at a relatively higher temperature ($\sim$
0.19~K). The change in thermal conductivity at this temperature is
just comparable with experimental resolution. As seen in the lower
panel, however, the electrical resistivity at T$_{c}$ has
increased by $\sim 20 \%$. Assuming that subkelvin thermal
conductivity does not depend on impurity concentration (which is
the case of cuprates at optimal doping level), this data suggests
that the change in T$_{c}$ and resistivity with time are mostly
due to an increase in the level of disorder and not a change in
the carrier concentration (which would have left a stronger
signature on the thermal conductivity). In order to minimize the
contamination of our study by this evolution, whenever we compare
electrical and thermal conductivities of a given sample, we use
the thermal and electrical data \emph{measured at the same time}.
Optimally-doped (OP) and overdoped (OD) samples were found to be
more stable and no significant change in their electrical and/or
thermal conductivities was detected after thermal cycling and
aging.

An accurate determination of the doping level, $p$, in the five
samples used in this study (and listed in Table I) is not easy. We
used the value of T$_c$ as the relevant parameter to estimate the
doping level. Ando \emph{et al.}\cite{Ando00} have reported that
in La-doped Bi-2201 the variation of T$_{c}$ with the doping level
is much faster than that found in other families of cuprates. The
usual ``Bell shape'' curve can be expressed as:
\begin{equation}
\frac{T_c}{T_c^{max}}=1-a(p-p_{opt})^2 \label{dome}
\end{equation}

While in the ``universal'' expression for the doping dependence of
the critical temperature,\cite{Presland91} a= 82.6, the data in
La-doped Bi-2201\cite{Ando00} yields a=275. We have used the
latter value, together with T$_c^{max}$=10.2K and $p_{opt}$ =0.17.
A supplementary check on the doping level of our samples was
performed by measuring the Hall coefficient, $R_H$ in two other
samples of the same batch with a T$_c$ and resistivity indicative
of optimal doping. As expected, the magnitude of the Hall
coefficient normalized by the volume of the elementary cell,
$R_He/V_0$, was found to be close to the expected value for an
optimally-doped cuprate.\cite{Bel04} Taking into account all these
considerations, the accuracy of the doping level determined for
each sample is estimated to lie in a margin of $\pm$ 0.01.

\section{Charge transport}
\begin{figure}
\includegraphics[width=0.9\linewidth,angle=0,clip]{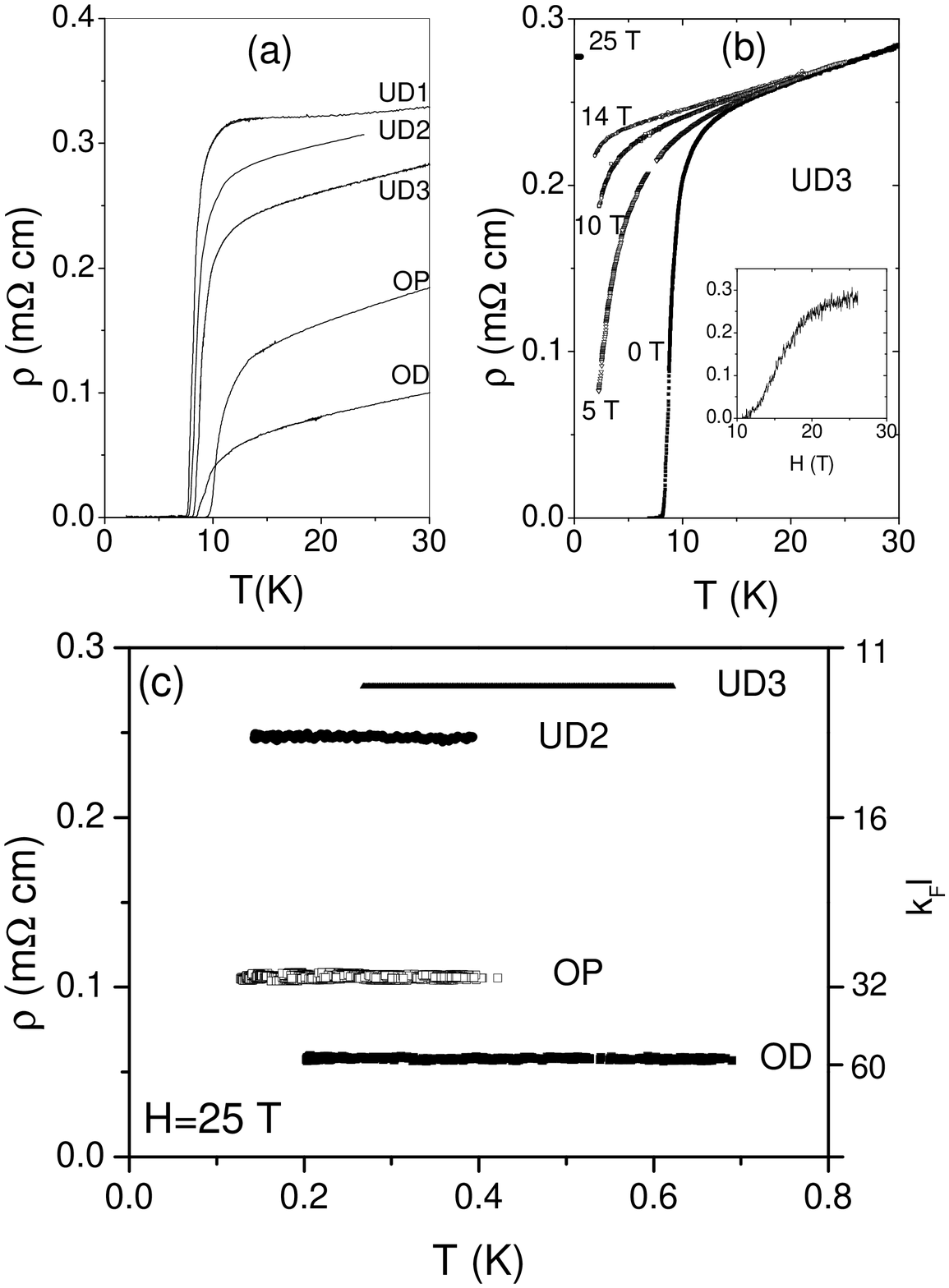}
\caption{ (a) Temperature dependence of the resistivity at zero
magnetic field. (b) Temperature dependence of the resistivity at
different magnetic fields for sample UD3. The inset shows a
magnetoresistivity curve obtained around 0.3~K (c) The temperature
dependence of resistivity for the four samples studied at B=25T.
Note that the resistivity is almost flat as a function of
temperature for all the samples investigated.} \label{fig3}
\end{figure}

Fig.~\ref{fig3}(a) shows the temperature dependence of the
resistivity of the five Bi-2201 samples. The magnitude of the
resistivity at T$_c$ varies from 0.05~m$\Omega$~cm for the
overdoped (OD) sample to 0.3~m$\Omega$~cm for the most underdoped
(UD) sample, in good agreement with recent investigation of charge
transport in Bi-2201 at different doping levels.\cite{Vedeneev04}
Note that 0.3~m$\Omega$~cm translates to the Ioffe-Regel parameter
$k_Fl\simeq11$, which corresponds to a metallic regime.
Fig.~\ref{fig3}(b) shows the resistivity versus temperature of
sample $UD3$ at different magnetic fields. The inset presents the
magnetoresistivity measured around 0.3~K which shows that
resistive transition is complete at H=25~T. As revealed by the
careful study in ref. \onlinecite{Vedeneev04}, the resistivity of
underdoped sample show an upward curvature in presence of a
magnetic field below T$_c$, but it saturates at very low
temperature. The same trend is observed in the two underdoped
samples. Fig.~\ref{fig3}(c) displays the temperature dependence of
the resistivity of four samples measured at 25~T for testing the
WF law. For samples UD2 and UD3, the value of the residual
resistivity at H=25~T is slightly higher than the value at T$_c$,
but saturates to a well-defined value below T=1~K. \footnote{Note
that the resistivity curve of sample UD2 in Fig.~\ref{fig3}(c)
corresponds to a first measurement whereas the curve in
Fig.~\ref{fig3}(a) corresponds to a second measurement and
displays a higher absolute resistivity.} This is an indication
that the doping level of the underdoped samples lies close to the
metal-insulator transition but it stays in the metallic side.


\section{Superconducting state}
Before addressing the test of WF law in Bi-2201, we begin by
presenting the thermal conductivity of the superconducting state,
in the absence of magnetic field.

\begin{figure}
\includegraphics[width=0.9\linewidth,angle=0,clip]{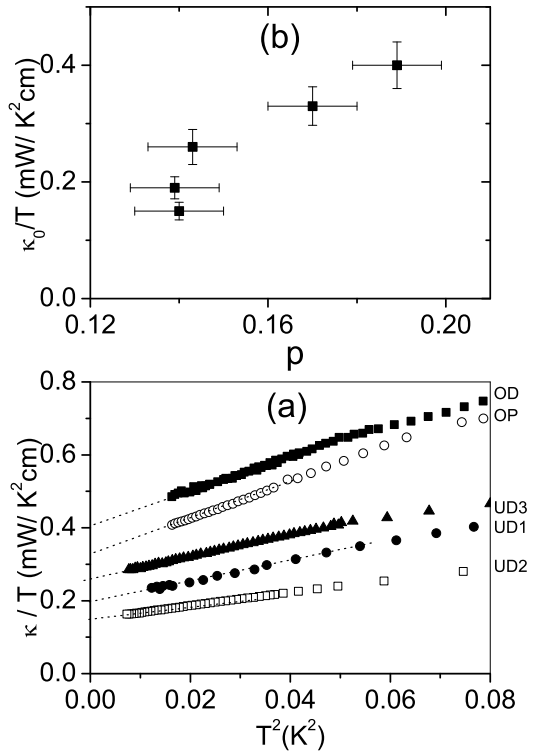}
\caption{ a): Thermal conductivity at low temperature for
different samples plotted as $\kappa$/T versus T$^2$. The dotted
lines represent linear extrapolations to T=0. b): Doping
dependence of the linear term of the thermal conductivity
extrapolated to T=0 (squares). } \label{fig4}
\end{figure}

The lower panel of Fig.~\ref{fig4} displays the temperature
dependence of thermal conductivity of the five Bi-2201 samples at
zero magnetic field. The data is plotted as $\kappa/T$ versus
$T^2$ in order to distinguish the phonon contribution to the
thermal conductivity. Indeed, at very low temperatures, when the
scattering of phonons becomes ballistic (e.g. limited by the
surface of the sample), the total thermal conductivity can be
written as:
\begin{equation}
\kappa(T)=aT+bT^3
 \label{e1}
\end{equation}
where the linear term is the electronic contribution. In a
$d$-wave superconductor, this corresponds to heat transport by
nodal quasiparticles. The second term represents the phonon
contribution. Hence, plotting $\kappa/T$ \emph{vs.} $T^2$, the
linear term can be obtained by extrapolating the data with a
straight line to the T=0 axis. The magnitude of the linear terms
$\kappa_0/T$ versus doping level is shown in the upper panel of
Fig.~\ref{fig4}. It can be noticed that $\kappa_0/T$ increases
when the doping level increases, in agreement with previous
studies on other families of cuprate superconductors
\cite{Takeya02,Sutherland03,Sun04,Ando04} and confirming the trend
already sketched in our previous communication.\cite{Bel04}
 At optimum doping level, we estimate
$\kappa_0/T$ to be 0.33~$mW/K^2 cm$. This is almost twice the
value reported in optimally-doped YBCO\cite{Chiao00} and
Bi2212\cite{Chiao00} with much higher T$_{c}$s and a considerably
larger superconducting gap. It is also twice larger than the value
($\sim 0.16~mW/K^2 cm$) reported for La-doped Bi-2201 at the same
doping level.\cite{Ando04} The difference between La-free and
La-doped Bi-2201 could be intrinsic and a consequence of the
higher T$_{c}$ (38~K compared to 10~K ) in the latter system.
However, one cannot exclude an experimental artefact. Indeed, an
anomalous downturn is present in the data reported for La-doped
Bi-2201.\cite{Ando04} In other words, below a certain
sample-dependent temperature, $\kappa/T$ suddenly appears to
decrease faster as a function of $T^2$. Such a downturn, known to
contaminate thermal conductivity measurements at very low
temperatures,\cite{Smith05} would lead to an underestimation of
$\kappa_{00}$ in La-doped Bi-2201.

\begin{table}
\caption{\label{table1} Physical characteristics of the Bi-2201
samples in this study: T$_c$, $p$ (number of carriers per Cu
atom), residual linear term in the thermal conductivity as well as
the gap maximum $\Delta_0$ deduced from Eq.~\ref{e4}.}
\begin{ruledtabular}
\begin{tabular}{|c|c|c|c|c|}
 sample&T$_c$&p&$\kappa_0$/T&$\Delta_0$ \\
 &(K)&&(mW/K$^2$ cm)&(meV) \\
 \hline
 OD&9.2&0.189$\pm0.01$&0.4$\pm0.04$&7.1$\pm0.7$ \\
 OP&10.2&0.17$\pm0.01$&0.33$\pm0.03$&8.6$\pm0.9$ \\
 UD1&7.5&0.139$\pm0.01$&0.19$\pm0.02$&14.9$\pm1.5$ \\
  UD2&7.7&0.14$\pm0.01$&0.15$\pm0.02$&18.8$\pm1.9$ \\
   UD3&8.1&0.143$\pm0.01$&0.26$\pm0.03$&10.8$\pm1.2$ \\
\end{tabular}
\end{ruledtabular}
\end{table}

BCS theory, when applied to a $d$-wave superconductor, associates
the linear term in thermal conductivity to the fine structure of
the superconducting gap at nodes. This association can be written
as \cite{Durst00}:
\begin{equation}
\frac{\kappa_0}{T} \simeq
\frac{k_B^2}{3\hbar}\frac{n}{d}\frac{v_F}{v_2}
 \label{e2}
\end{equation}
Here $v_F$ and $v_2$ are the velocities of nodal quasiparticles
normal and parallel to the Fermi surface, respectively, and $n/d$
is the number of $CuO_2$ planes per unit cell. In the clean limit,
when $\hbar\gamma \ll k_BT_c$, this linear term is $universal$ in
the sense that it does not depend on the scattering time. Assuming
a pure $d_{x^2-y^2}$ symmetry for the superconducting gap and
given that
\begin{equation}
v_2=\frac{1}{\hbar k_F} \left. \frac{d\Delta}{d\phi}
\right|_{node}
 \label{e3}
\end{equation}
the magnitude of the superconducting gap is directly related to
the magnitude of the linear term
\begin{equation}
\Delta_0=\frac{k_B^2}{6} \frac{n}{d} \frac{k_F v_F}{\kappa_0/T}
 \label{e4}
\end{equation}

Using the typical Fermi surface parameter for
Bi$_2$Sr$_2$CaCu$_2$O$_8$ (Bi-2212) namely k$_F$=0.7~\AA  along
the nodal direction and v$_F=2.5\times10^7$~cm/s, and $n/d=2/c$
with $c=24.6~\AA$ for Bi-2201, we obtain $\Delta_0 \simeq
8.6\pm0.9~meV$ at optimal doping \footnote{Note that the value
calculated in ref.~\onlinecite{Bel04} is overestimated by a factor
2}. It should be compared to the value for other cuprates at
optimum doping estimated using the same method: namely, 30~meV for
Bi-2212 and 50~meV for YBCO. It is noteworthy that while T$_c$ of
La-free Bi-2201 is reduced by almost one order of magnitude
compared to Bi-2212 the magnitude of the gap is reduced by a much
smaller factor. It is instructive to compare the value deduced
from thermal conductivity to what has been directly measured using
tunnelling ($\Delta_{tunnel} \simeq$ 12-15 meV
\cite{Kugler01,Vedeneev01}) and ARPES ($\Delta_{ARPES} \simeq
10\pm2~meV$\cite{Harris97}).

One should not forget that in Bi-2201, the impurity bandwidth
$\hbar\gamma$ is not negligible compared to $k_BT_c$.  Indeed,
using the Drude formalism for resistivity, $\rho=m/ne^2\tau$, and
the plasma frequency, $\omega_p^2=ne^2/\epsilon_0m$ and given that
$\rho(T_c) \simeq 100~\mu \Omega~cm$, $\hbar\omega_p=8300~cm^{-1}$
at optimal doping,\cite{Romero92} we can estimate the impurity
bandwidth in the unitary limit
$\hbar\gamma=0.63\sqrt{\Delta_0/2\tau} \simeq 5~meV$, which is a
sizeable fraction of the superconducting gap. Since any correction
to the universality should increase the magnitude of the linear
term,\cite{Sun95} the intrinsic linear term (i.e. not renormalized
by disorder) is actually smaller than the measured one and
therefore the actual magnitude of the superconducting gap maybe
higher than 8.6~meV.

It is instructive to compare our Bi-2201 samples to highly
overdoped Tl$_2$Ba$_2$CuO$_6$ \cite{Proust02} with $T_c\simeq
15~K$, for which the linear term in the superconducting state is
1.41 mW/K$^2$~cm and translates to  $\Delta_0\simeq2.1~meV$,
comparable to the BCS prediction $2.14k_BT_c\simeq2.8 meV$. In
Bi-2201, around optimal doping, the magnitude of the
superconducting gap deduced from thermal conductivity is about
five times larger than the BCS prediction. This difference has
been interpreted by assuming that the gap seen in thermal
conductivity is related to the pseudogap.\cite{Sutherland03} In
the d-density-wave (DDW) scenario for example, the pseudogap has
$d_{x^2-y^2}$ symmetry with linear dispersion near the node.
Thermal conductivity is universal provided that the chemical
potential is neglected and the nesting is perfect. The linear term
probes thus a gap which is not purely superconducting but can be
expressed as $\Delta=\sqrt{\Delta_1^2+\Delta_2^2}$ where
$\Delta_1$ and $\Delta_2$ are the order parameters of $DDW$ and
$d$-wave superconductivity, respectively.\\

\section{Test of the Wiedemann-Franz law}

\begin{figure}
\includegraphics[width=0.9\linewidth,angle=0,clip]{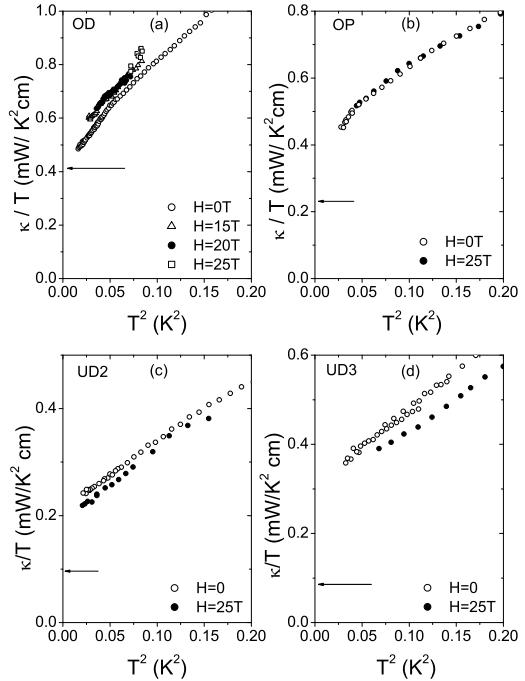}
\caption{Comparison of thermal conductivity in the superconducting
state (open symbols) and in the normal state (solid symbols) for
(a) an overdoped sample, (b) an optimally-doped sample, (c) and
(d) two slightly underdoped samples. Note the increase of
$\kappa$(H) seen in the overdoped sample and the decrease of
$\kappa$(H) in the underdoped samples. Arrows mark the expected
value for $\kappa$/T in the normal state according to the WF law.
The zero-temperature extrapolation becomes significantly larger
than this value in the underdoped samples (see text).}
\label{fig5}
\end{figure}

Let us now turn our attention to the verification of the WF law. A
comparison of Subkelvin thermal conductivity at zero field and
25~T is shown in Fig.~\ref{fig5} for four samples which were
measured at high magnetic field. Note that this data correspond to
measurements performed on \emph{as-grown} samples. For sample OD
(Fig.~\ref{fig5}a), the application of the magnetic field leads to
a slight, yet visible increase in thermal conductivity. At optimal
doping (Fig.~\ref{fig5}b), the thermal conductivity is almost
identical for normal and superconducting states. However, for the
underdoped samples UD2 (Fig.~\ref{fig5}c) and UD3
(Fig.~\ref{fig5}d), thermal conductivity decreases with the
application of magnetic field as it does in underdoped
LSCO.\cite{Takeya02,Hawthorn03} In the latter case, the
field-induced decrease in thermal conductivity was interpreted as
a field-induced thermal metal-insulator transition presumably as a
consequence of a competing order with superconductivity such as
spin-density-wave\cite{Gusynin04}. The existence of such competing
order has been theoretically predicted in the phase diagram of
cuprates when disorder plays a prominent role\cite{Chen04}.
Sutherland \emph{et al.}\cite{Sutherland05} report that, in
contrast to LSCO, underdoped YBCO , for a doping level of $\delta
\approx 0.33$ (corresponding to the onset of superconductivity),
is a thermal metal. In other words, it does not display a
detectable field-induced decrease in thermal conductivity. This
conclusion is not shared by Sun \emph{et al.} \cite{Sun04,Sun05}
who report a metal-insulator transition around $\delta=0.45$. The
level of disorder in Bi-2201 is certainly closer to LSCO than to
ultra-clean YBCO \cite{Liang98}. Therefore, it is not surprising
that, underdoped Bi-2201 displays a field-induced decrease in
$\kappa_{0}/T$ in analogy with LSCO.

\begin{figure}
\includegraphics[width=0.9\linewidth,angle=0,clip]{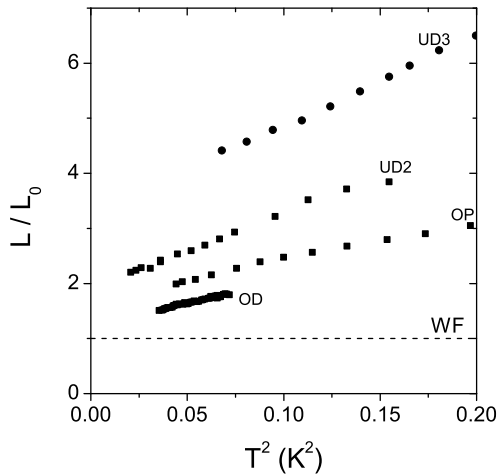}
\caption{Lorentz number normalized by the Sommerfeld value L$_0$
versus T$^2$ for different samples at various doping level in the
normal state. Dashed line is the WF expectation at T=0.}
\label{fig6}
\end{figure}

A second feature of the data presented in Fig.~\ref{fig5} is the
upward deviation of any zero-temperature extrapolation of the
normal-state thermal conductivity compared to what is expected
according to the WF law. As seen in the figure, such a deviation
becomes striking in the underdoped samples. In Fig.~\ref{fig6}, we
have plotted the Lorenz number $L=\kappa\rho_0/T$ normalized by
$L_0$ versus T$^2$, where $\rho_0$ is the normal state resistivity
(measured at 25T, using the data of Fig.~\ref{fig3}c). Dashed line
shows the value expected according to the WF law. For the OD
sample (p=0.19), we recover the WF law within a margin 5 $\%$
\footnote{ For this sample, we had only a  few data points  for
H=25~T. Since they indicated that $\kappa/T$ becomes
field-independent above 15~T, we used the H=20~T data for
estimating L$_{0}$.  Note that the margin of verification in this
sample is somewhat smaller than what was reported in
ref.~\onlinecite{Bel04}}. This result is in agreement with the
behavior observed for  overdoped Tl-2201(p=0.26).\cite{Proust02}
For the OP sample, a magnetic field of 25~T leaves thermal
conductivity unchanged for the explored temperature range. Thus,
we have used the zero-field extrapolation as the residual normal
state thermal conductivity, which gives a Lorenz ratio L=1.3L$_0$,
significantly larger our uncertainty. It can be noticed that the
deviation of the WF law when T$\rightarrow$ 0 becomes more
pronounced for underdoped samples. Indeed, the T=0 extrapolation
of the thermal conductivity in the normal state of sample UD2
(p=0.14) leads to a Lorenz ratio L=1.9L$_0$. For sample
UD3(p=0.143),  the extrapolation to T=0 leads to a violation of
the WF law by a factor of 3. However, given the lack of low
temperature data for this sample, this value could be
overestimated.

\section{Discussion}

\begin{figure}
\includegraphics[width=0.8\linewidth,angle=0,clip]{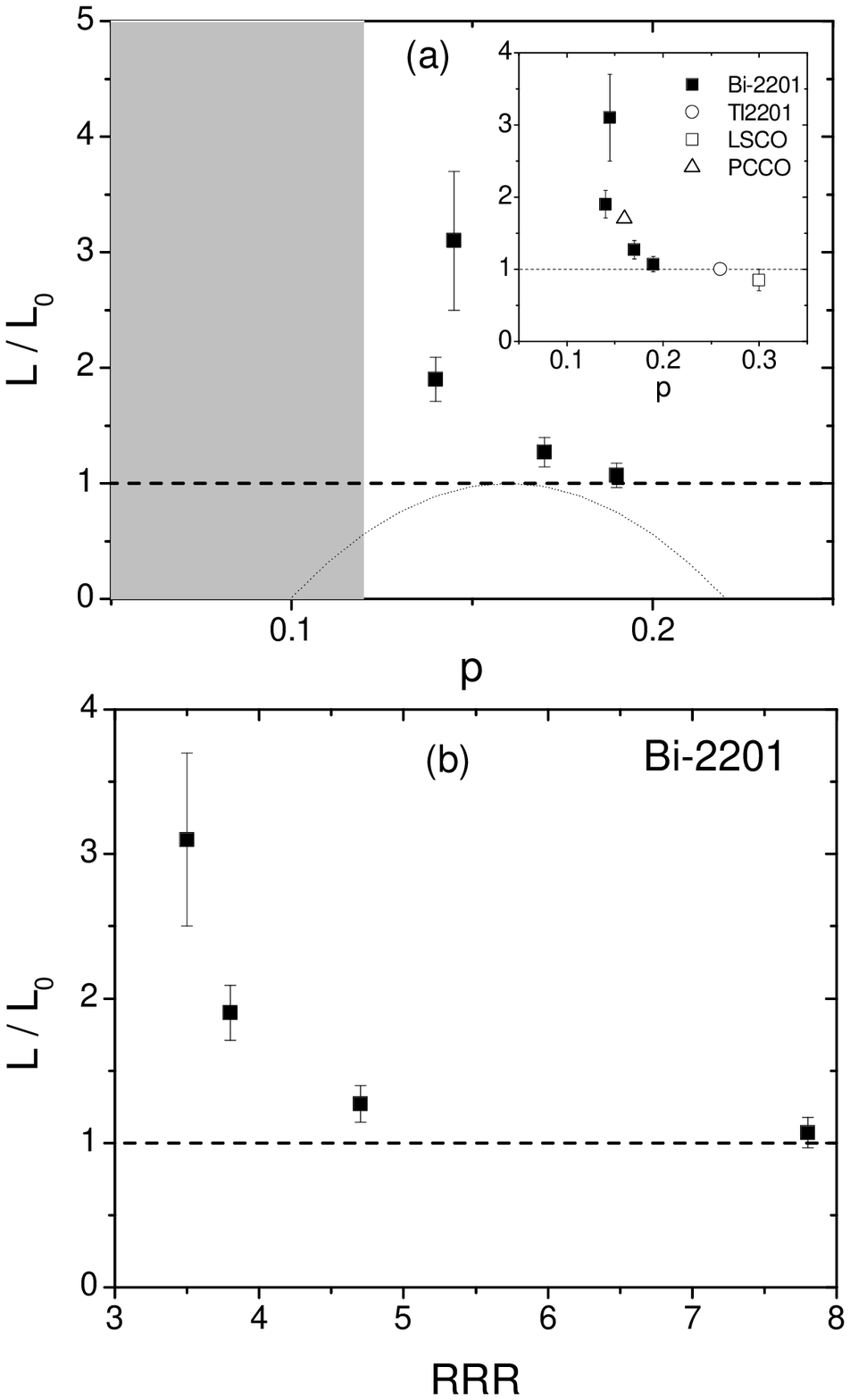}
\caption{(a) The main panel shows the doping dependence of the
Lorentz ratio normalized by L$_0$ for all Bi-2201 samples (full
squares). The inset is a comparison with data reported in previous
studies for PCCO,\cite{Hill01} Tl-2201,\cite{Proust02} and
LSCO.\cite{Nakamae03} The horizontal line is the WF expectation.
The shaded area marks the cross-over from metal-to-insulator in
La-doped Bi-2201.\cite{Ono00} (b) Lorentz ratio normalized by
L$_0$ versus residual resistivity ratio for Bi-2201 (see text).}
\label{fig7}
\end{figure}

In the main panel of Fig.~\ref{fig7}(a), the Lorenz ratio
normalized by L$_0$ is plotted versus carrier concentration for
all Bi-2201 samples in this study. There is also a sketch of the
superconducting dome given by Eq.~\ref{dome} with the La-doped
Bi-2201 parameter. In the inset of Fig.~\ref{fig7}(a), we
summarized the available data on the validity of the WF law in
cuprate superconductors. In addition to our Bi-2201 data (full
squares), we put data on LSCO \cite{Nakamae03} (open square),
Tl-2201 \cite{Proust02} (open circle) and PCCO \cite{Hill01} (open
triangle).

These results indicate that the WF law is verified in the
overdoped side of the phase diagram. Indeed, it has been now
verified in three families of cuprate superconductors. A departure
starts to develop at optimal doping. It becomes particularly
pronounced at a doping level (p $\sim$ 0.14) lying close to the
metal-to-insulator cross-over. The shaded area represents the
``insulating'' ground state associated with a logarithmic
divergency of the resistivity, which is reported to occur for
$p\leq 0.12$ both in La-doped\cite{Ono00} and in La-Free
Bi-2201.\cite{Vedeneev04} Obviously, the WF law is not relevant
for an insulator lacking delocalized fermions. However, as seen in
Fig. ~\ref{fig3}(c), the samples presented in this study are not
insulators and their resistivity display almost no temperature
dependence below 1K.

According to our current understanding, the observed violation of
the WF law in underdoped Bi-2201 samples cannot be attributed to
any known experimental artefact. The magnitude of the departure
($\frac{L}{L_{0}}\geq$ 2) is well above our experimental
resolution. It is reproducible and becomes systematically larger
with the increase in disorder and/or underdoping. While the
underdoped samples may host an inhomogeneous distribution of
oxygen content as discussed above, this is expected to affect the
transport of heat and charge in the same way. In the absence of
any plausible scenario linking an excess of heat conductivity to
extrinsic effects or to a macroscopic inhomogeneity, we should
consider the possible microscopic origins of such a violation.

The first class of available scenarios are based on spin-charge
separation. A strong violation of the WF law is predicted in the
case of a Luttinger liquid.\cite{Kane96} Indeed, in a
one-dimensional interacting electron system, quasi-particles are
replaced by collective excitations of charge (without spin) and
spin (without charge) moving independently and at different
velocities. Since the electrical current probes only the charge
excitations, while the thermal current is sensitive to both charge
and spin excitations, there is an excess in thermal conductivity
compared to electric conductivity. An upward deviation from the WF
law is thus expected. A similar violation of the WF law is
expected in theories invoking electron fractionalization in
cuprates.\cite{Senthil01} We note, however, that other
experimental signatures expected in case of such a
fractionalization have not been detected.\cite{Bonn01}

Another class of models are those invoking the breakdown of the
Fermi liquid in the vicinity of a Quantum Critical Point(QCP). The
existence of a hidden order such as a d-density-wave
\cite{Chakravarty01}has been proposed in order to explain the
pseudogap phenomena. Analysis of the transport properties of such
a state\cite{Kim02,Sharapov03} indicates that the WF law should
remain intact in the zero-temperature limit. Close to a QCP,
however, one may expect the emergence of non-Fermi liquid
properties\cite{Custers03} which may include a violation of the WF
law. We note, however, that a study of the WF law in
CeNi$_{2}$Ge$_{2}$, a heavy-fermion compound lying close to a QCP,
did not find any detectable departure from the WF
law.\cite{Kambe99}

Until very recently, it was widely believed that the WF law is a
robust property of any Fermi liquid. In particular, Castellani and
co-workers \cite{Castellani87} established the validity of the WF
law in interacting disordered electron system up to the
metal-insulator transition. However, this issue has been
reexamined recently by three independent
groups\cite{Raimondi04,Niven05,Catelani04} who all suggest the
violation of the WF law in a disordered conductor with interacting
electrons. Roughly speaking, this violation arises because the
Coulomb interaction leads to an additional scattering at low
temperatures which impedes charge transport more efficiently. In
such a context, the expected deviation from the WF law is positive
and  scales with 1/k$_F\ell$.\cite{Raimondi04} Therefore, for the
most underdoped Bi-2201 sample, the expected deviation is
$\frac{L}{L_{0}}\sim1.1$. Such a deviation is an order of
magnitude smaller than observed
experimentally$\frac{L}{L_{0}}\sim2$. This quantitative
disagreement should however be put into context. The origin of the
logarithmic divergence of resistivity in cuprates which occurs for
surprisingly high values of k$_F\ell$ is still far from
understood. If this last line of speculation happens to be the
relevant one, then disorder, more than doping level, would be the
key parameter. Fig.~\ref{fig7}(b) presents the Lorenz ratio
normalized by L$_0$ versus the residual resistivity ratio (RRR),
that is the ratio between the room temperature resistivity and the
normal state resistivity at H=25~T and T$\rightarrow$ 0. The
latter is a common parameter for quantifying the level of
disorder. As seen in the figure, the magnitude of L/L$_{0}$ is in
excellent correlation with the decrease in RRR. This correlation
appears to be even more robust than the one observed between
L/L$_{0}$ and the doping level displayed in the upper panel.

\section{Conclusion}
We have studied low-temperature thermal transport in several
crystals of Bi-2201 with different doping levels. In the
superconducting state, the residual linear term of thermal
conductivity  steadily increases with the increase in the doping
level as previously reported for other families of the cuprates.
The application of the magnetic field leads to an \emph{increase}
in thermal conductivity in overdoped regime and a \emph{decrease}
in the underdoped regime. In the normal state, we have tested the
Wiedemann-Franz law and confirmed the validity of this fundamental
law in overdoped regime. A departure from the WF law appears at
optimal doping and becomes more pronounced for underdoped samples.
This provides experimental evidence for the violation of the WF
law in the vicinity of metal-insulator transition in cuprates.

\section{Acknowledgement}

We thank P. van der Linden for technical assistance and L.
Taillefer, N. E. Hussey and P. Schwab for useful discussions.

\bibliography{kappa1}

\end{document}